\documentclass[9pt,twocolumn,twoside]{osajnl}

\journal{ol} 
\usepackage{academicons}
\usepackage{hyperref}
\usepackage{xcolor}
\usepackage{graphicx}
\usepackage{lipsum}
\setboolean{shortarticle}{true}

\title{Inverse design of Raman amplifier in frequency and distance domain using Convolutional Neural Networks }

\author[1,*]{Mehran Soltani \href{https://orcid.org/0000-0002-5831-1296}{\includegraphics[scale=1]{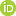}}}

\author[1]{Francesco Da Ros \href{http://orcid.org/0000-0002-9068-9125}{\includegraphics[scale=1]{figures/ORCIDiD_icon16x16.png}}}
\author[2]{Andrea Carena \href{http://orcid.org/0000-0001-6848-3326}{\includegraphics[scale=1]{figures/ORCIDiD_icon16x16.png}}}
\author[1]{Darko Zibar \href{http://orcid.org/0000-0003-4182-7488}{\includegraphics[scale=1]{figures/ORCIDiD_icon16x16.png}}}

\affil[1]{DTU Fotonik, Technical University of Denmark (DTU), DK-2800 Kgs. Lyngby, Denmark}
\affil[2]{Dipartimento di Elettronica e Telecomunicazioni (DET), Politecnico di Torino, Corso Duca degli Abruzzi, 24 - 10129, Torino, Italy}

\affil[*]{Corresponding author: msolt@fotonik.dtu.dk}

\begin{abstract}
We present a Convolutional Neural Network (CNN) architecture for inverse Raman amplifier design. This model aims at finding the pump powers and wavelengths required for a target signal power evolution, both in distance along the fiber and in frequency. Using the proposed framework, the prediction of the pump configuration required to achieve a target power profile is demonstrated numerically with high accuracy in C-band considering both counter-propagating and bidirectional pumping schemes. For a distributed Raman amplifier based on a 100 km single-mode fiber, a low mean set (0.51, 0.54 and 0.64 dB) and standard deviation set (0.62, 0.43 and 0.38 dB) of the maximum test error are obtained numerically employing 2 and 3 counter, and 4 bidirectional propagating pumps, respectively. 
 
\end{abstract}

\setboolean{displaycopyright}{true}

\begin{document}

\maketitle

\section{Introduction}

In long-haul optical communications, amplifiers play a crucial role in compensation of the link losses. Designing a desired optical amplification scheme is challenging as the requirements in noise figure (NF) and gain profile can be strict. Regarding this, Erbium-doped fiber amplifiers (EDFAs) and distributed Raman amplifiers (DRAs) have been extensively researched. EDFAs are more power efficient while DRAs provide low NF. Moreover, DRAs' power profile can be adjusted easily by changing the power and wavelength of the pumps \cite{Agrawal, Desurvire}, which makes them more attractive for wideband wavelength division multiplexed (WDM) scenarios \cite{deMoura:20}.

One of the main challenges with inverse DRA design is to realize the pump configuration based on a desired gain at the end of the link. Several machine learning solutions have been proposed for this problem in the literature. In \cite{8894395}, a neural network (NN) model averaging along with a fine tuning technique is employed to learn the relationship between the desired gain profile at the end of fiber and the corresponding pump powers and wavelengths in a SMF link. Furthermore, \cite{9244561} proposes an Autoencoder (AE) scheme by embedding a differentiable Raman amplifier model in the training procedure of a NN to predict the pump parameters for a specific family of gains in a FMF link.  


Alternatively to designing the power spectral density at the amplifier output (frequency domain), being able to control the power evolution over the transmission span (spatial domain) offers a variety of advantages. Uniform distribution of the power along the span results in a quasi-lossless transmission which minimizes the amplified spontaneous emission (ASE) noise level \cite{Ania-Castanon:04,Quasi-lossless1, 1601058}. Such a power distribution would also help several of the Kerr nonlinearity mitigation techniques currently being investigated, e.g. transmission based on the nonlinear Fourier transform theory which assumes lossless transmission \cite{le2015nonlinear} or nonlinearity mitigation using mid-link optical phase conjugation which requires a symmetric power distribution \cite{tan2018distributed}. A significant research effort has been devoted both numerically and experimentally into achieving a desired signal power evolution over a narrow frequency bandwidth but with limited work presented on full C-band. So far, no optimization method has been proposed for addressing the power evolution design jointly in spatial and frequency domain. 

 In this paper, we propose a supervised deep CNN architecture for inverse DRA design in a SMF link to find the pump powers and wavelengths based on the two dimensional signal power profile in frequency and distance along the fiber. The proposed method employs two networks trained in an end-to-end procedure: A CNN network as the feature extraction followed by a multi-layer NN as the regression for predicting the pump power and wavelength values based on the extracted features. The proposed method reduces the high spatial redundancy of the signal power in frequency and distance and also extracts informative features for prediction of the pump parameters. For the evaluation of this framework, numerical simulations are demonstrated in C-band utilizing both counter-propagating and bidirectional pumping schemes. 

The remainder of the paper is organized as follows. In section II, systems level CNN-based architecture and the training configuration for inverse DRA design is described. Section III demonstrates the numerical simulation results of the proposed method for counter and bidirectional propagating schemes in C-band. Finally, section IV concludes the paper.

\begin{figure*}
\centering

\includegraphics[width=\textwidth,height=4.5cm]{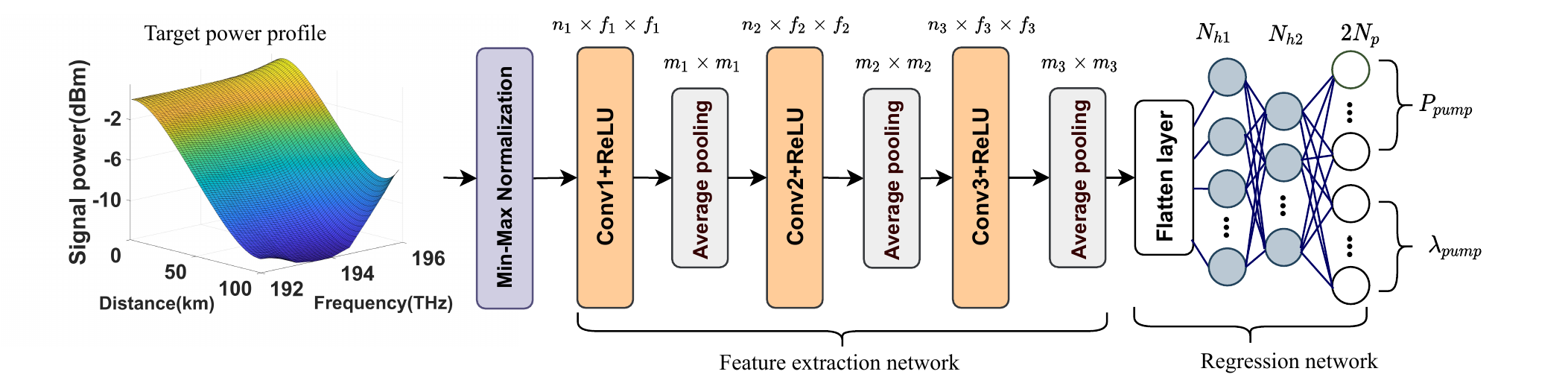}

\caption{Diagram of the CNN architecture for the inverse DRA design based on target power profile in frequency and distance}
\label{fig:Raman_Net}
\end{figure*}

\section{System level inverse DRA design }




\subsection{CNN architecture for inverse DRA design}
Considering a system with a forward mapping denoted as $Y = f(X)$, the inverse design problem aims at modeling the mapping $f^{-1}(.)$ in order to find the input $X$ providing a desired target output $Y$. In many cases, the forward mapping is straightforward and can be solved numerically, while the inverse mapping may be very complex or even unknown. Regarding this, machine learning methods, especially deep learning algorithms, have shown a promising performance in learning the approximate inverse mapping between $X$ and $Y$ only based on samples $(X^s,Y^s)$ generated from the forward model $f(.)$ \cite{8894395}. For simplicity in our further analysis, we will refer to $X$ and $Y$ as the sampled versions of the input and output spaces.

In inverse DRA design for a SMF link based on signal power evolution, the forward mapping can be described as $\textbf{P}_s(f,z) = f([\textbf{P}_{pump};\mathbold{\lambda}_{pump} ])$  where $\textbf{P}_s(f,z) = [p_{ij}]_{N_{ch}\times N_z}$ is the two dimensional signal power in which $p_{ij}$ is signal power at $i$-th frequency channel and $j$-th distance index in a WDM system with $N_{ch}$ number of channels and $N_z$ distance points,  $f(.)$ is a system of nonlinear differential equations for Raman amplification scheme \cite{Agrawal}, $\textbf{P}_{pump}=[P_1,\cdots, P_{N_p}]^T$ is the pump power vector with $T$ denoting the transpose operator, and $\mathbold{\lambda}_{pump}=[\lambda_1,\cdots, \lambda_{N_p}]^T$ is the pump wavelength vector.

 To be able to simultaneously predict the pump powers and the wavelengths, we present a deep learning algorithm to model the inverse mapping $[\textbf{P}_{pump};\mathbold{\lambda}_{pump} ]=f^{-1}(\textbf{P}_s(f,z))$. The dimensionality of the problem does not allow to simply apply the frameworks of \cite{8894395, 9244561}, since in order to make the input compatible with these methods, the power profile for each sample should be converted to an array of length ${N_{ch}\times N_z}$ which results in an extremely large and complex network. For instance, if a system have $N_{ch} = 40$ channels and 100 km span length with distance resolution of 1 km, $N_z=100$, the number of the nodes of the input layer will be $N_{ch}\times N_z=4000$. The mapping between such a high dimensional input and the pumping configuration requires a network with high number of trainable parameters which not only takes too much time to be trained, but also will be prone to problems like overfitting and local minimum. Additionally, using the approach of \cite{8894395, 9244561} would be unnecessarily complex as it would not take advantage of the inherent correlation between the input data. Clarifying this, each point in a WDM frequency-distance space resembled as a pixel of a two dimensional image, has a high spatial redundancy. This means that adjacent points have a lot amount of information in common, however, fully-connected NNs are not capable of reducing these spatial redundancies. Concerning these two main problems, we found CNNs more attractive since they have been designed to process data coming on the form of multiple arrays, like images, and moreover, they can successfully capture the Spatial and Temporal dependencies in a two dimensional form data through the application of relevant filters \cite{CNN} and weight sharing. 

For a CNN-based demonstration of inverse DRA design, we consider the distance-frequency power evolution matrix $\textbf{P}_s(f,z)$ as a two dimensional input to the network aiming to predict the pump configuration leading to the target $\textbf{P}_s(f,z)$. Diagram of the CNN-based method is illustrated in Fig.\ref{fig:Raman_Net}. The proposed framework is made up of two stages trained End-to-End, a \textit{feature extraction},
\begin{equation}
R(.;\theta_R): \mathbb{R}^{N_{ch}\times N_z}\rightarrow \mathbb{R}^{q\times r\times n_3}, q\times r\times n_3< N_{ch}\times N_z
\label{eq:CNN_model2}
\end{equation}
and a \textit{regression network}, 
\begin{equation}
F(.;\theta_F): \mathbb{R}^{(q\times r\times n_3)\times 1}\rightarrow \mathbb{R}^{2N_p\times 1}
\label{eq:CNN_model3}
\end{equation}
with trainable parameters $\theta_R$ and $\theta_F$, respectively.
First, a pixel-wise min-max normalization is performed on the input data as a pre-processing step. The minimum and maximum values selected for each frequency-distant points are equal to the the minimum and maximum values of the points in the training set, respectively. Afterwards, the normalized profile is passed to the feature extraction network $R(.;\theta_R)$ which consists of three CNN layers with $n_1$, $n_2$ and $n_3$ filters of size $f_1\times f_1$, $f_2\times f_2$ and $f_3\times f_3$, respectively. Moreover, each CNN layer is followed by rectified linear unit ($ReLU(x)=max(0,x)$)  as the activation function which can speed up the training process due to its simplicity in gradient calculation. Furthermore, spatial pooling is carried out by three average-pooling layers inserted in between successive CNN layers with the window size of $m_1\times m_1$, $m_2\times m_2$ and $m_3\times m_3$, respectively. The function of pooling layers is to progressively reduce the spatial size of the input feature maps resulting in lower amount of parameters and computations in the network. It is worth noting that the reduction of the scale of representation by each pooling layer is equal to its window size. Consequently, each layer of this network generates informative and compact representations of the input through nonlinear mappings. The output of the last CNN layer is a three-dimensional representation of the input profile consisting of $n_3$ different two-dimensional representations generated by the different filters of the last layer each with the spatial sizes of $q={N_{ch}}/({m_1\times m_2\times m_3})$ and $r={N_z}/({m_1\times m_2\times m_3})$. This 3D-representation is converted thereafter to a vector of length $q\times r\times n_3$ by the flatten layer and then is passed to the regression network $F(.;\theta_F)$. The objective of this network, modeled as a deep fully-connected, is to map the extracted features to the pumping setup. This network has four layers including the flatten layer of size $q\times r\times n_3$, two hidden layers of size $N_{h1}$ and $N_{h2}$ and the last layer of size $2N_p$, representing the pumping configuration vector. The values of $N_{h1}$ and $N_{h2}$ are optimized depending on the proposed pump configuration. 
 
\subsection{Training and evaluation}

Since the proposed approach relies on supervised learning, a data-set $D=\{Y_k,X_k|k=1,\cdots, K\}$ needs to be generated where $K$ is the number of samples, $Y_k=[\textbf{P}_{k_{pump}};\mathbold{\lambda}_{k_{pump}}]$ and $X_k = \textbf{P}_{k_s}$ are the pumping configuration vector and the corresponding 2D signal profile of the $k$-th sample, respectively. In this paper we focus on a data-set generated by solving the Raman amplifier differential equations \cite{Agrawal}, denoted as \textit{Raman solver}, for different pump powers and wavelengths. For each sample data, each value of the pump parameters denoted as the $m$-th value of the vector $Y$ have been selected based on a uniform distribution
\begin{equation}
y_m \sim \textit{U}[y_m^{min}, y_m^{max}]
\label{eq:cost function}
\end{equation}
in which $y_m^{min}$ and $y_m^{max}$ are the minimum and maximum values allowed to be taken by the $m$-th value of $Y$, respectively. After the data generation, same as the most supervised learning approaches, we divide the data into separate training, testing and validation sets. Also, we make sure that the training set contains the minimum and the maximum values of each dimensions of the input signal power evolution matrix $X_k$ to have a good generalization property \cite{8894395}. The overall model of the inverse design network can be described as:
\begin{equation}
 Y = R(F(X; \theta_F);\theta_R)
\label{eq:CNN_model1}
\end{equation}
in which both $R$ and $F$ are jointly trained to minimize the average cost function $C$ between the original target value $Y_l$ and the approximated value $\hat{Y_l}=R(F(X_l; \theta_F);\theta_R)$ of the training set:
\begin{equation}
\theta_R^*, \theta_F^*=\underset{\theta_R, \theta_F}{\operatorname{argmin}}\frac{1}{L}\sum_{l=1}^{L}C(\hat{Y_l},Y_l)
\label{eq:CNN_model}
\end{equation}
where $L$ is the number of training samples and for each sample, $C$ is the mean square error (MSE) value between the target and the approximated pump set-up values $Y_l$ and $\hat{Y_l}$, respectively:
\begin{equation}
C=\frac{1}{2N_p}\sum_{i=1}^{2N_p}(Y_l^{i}-\hat{Y_l}^{i})^2
\label{eq:cost function}
\end{equation}

The parameters of the network are updated in an iterative approach by means of gradient descent algorithm and back-propagation \cite{CNN}. Furthermore, advanced optimization algorithm RMSprop \cite{Tieleman2012} is employed for updating the parameters as it provides a fast and robust convergence for each parameter.

\begin{figure}
\centering
\includegraphics[width=8.5cm]{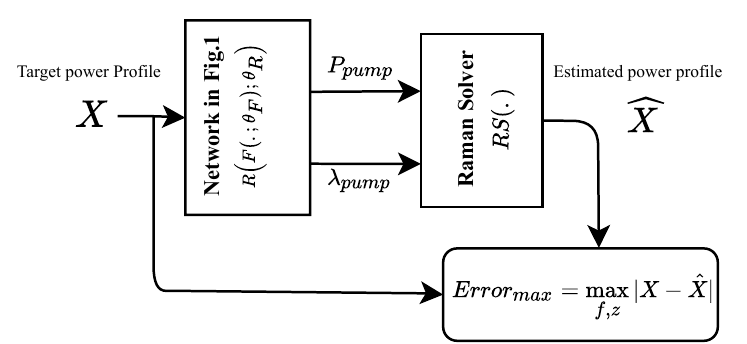}
\caption{Evaluation diagram of the proposed method}
\label{fig:Evaluation}
\end{figure}

Once the training of the network has been completed, we fix the set of the learnt parameters of the network $\theta=\{\theta_R,\theta_F\}$ to evaluate its performance. To this end, we put the network into a schematic as illustrated in Fig. \ref{fig:Evaluation}. In this scheme, for each input power profile, the corresponding pump powers and wavelengths have been predicted using the network in Fig. \ref{fig:Raman_Net} and then passed to the Raman solver $RS(.)$ to compute the power profile based on the predicted pumping setup. Afterwards, maximum absolute difference between the predicted and the input power profile is calculated in frequency (f) and distance (z) domain as the final prediction error for each sample. 

\section{Simulation results}

In this section, we investigate the CNN-based framework presented in the previous section for the design of Raman amplifiers in C-band. The data-set are generated using the Raman Solver provided by GNPy \cite{Gnpy}, an open source application developed recently for analyzing optical networks.

We consider a single span and analyze the evolution of the power profile jointly over the distance and the entire C-band (between 192 and 196 THz). Also, three propagation cases are deployed for the evaluation of the proposed method: two counter-propagating cases with 2 and 3 pumps and a bidirectional propagating case with 4 pumps (2co+2counter). The ranges for pump powers and wavelengths are specified in Table \ref{tab:pump parameters}. The superscripts (-) or (+) on the power ranges specify the counter or co-propagation of the corresponding pump, respectively.

\begin{table}[htbp]
\centering
\caption{\bf Power and wavelenght ranges for each DRA case}
\begin{tabular}{| m{1.2cm} | m{1.8cm}| m{2.3cm} |m{1.8cm} |}
\hline
Param & 2 pumps & 3 pumps & 4 pumps \\
\hline
$P_1[mW]$ & $[40-400]^-$ & $[30-300]^-$ & $[30-300]^-$ \\
$P_2[mW]$ & $[40-400]^-$ & $[30-300]^-$ & $[30-300]^-$ \\
$P_3[mW]$ &  - & $[30-300]^-$ & $[30-300]^+$ \\
$P_4[mW]$ & - & - & $[30-300]^+$ \\
$\lambda_1[nm]$ & $[1414-1449]$ & $[1414-1437.3]$ & $[1414-1449]$ \\
$\lambda_2[nm]$ & $[1449-1484]$ & $[1437.3-1460.3]$ & $[1449-1484]$ \\
$\lambda_3[nm]$ & - & $[1460.3-1484]$ & $[1414-1449]$ \\
$\lambda_4[nm]$ & - & - & $[1449-1484]$ \\
\hline
\end{tabular}
  \label{tab:pump parameters}
\end{table}

We divided the C-band into 40 channels with 100 GHz spacing. Input signal power per-channel is set to 0 dBm which results in a total WDM signal power of 16 dBm. Furthermore, a standard silica fiber with the following parameters is assumed: span length $L_{span}=$100 $km$, signal data attenuation $\alpha_s=0.2$ $dB/km$, pump power attenuation $\alpha_p=0.25$ $dB/km$, effective area $A_{eff}=80$ $\mu m^2$, non-linear coefficient $\gamma=1.26$ $1/W/km$.  

\begin{figure}
\centering
\includegraphics[width=8.5cm]{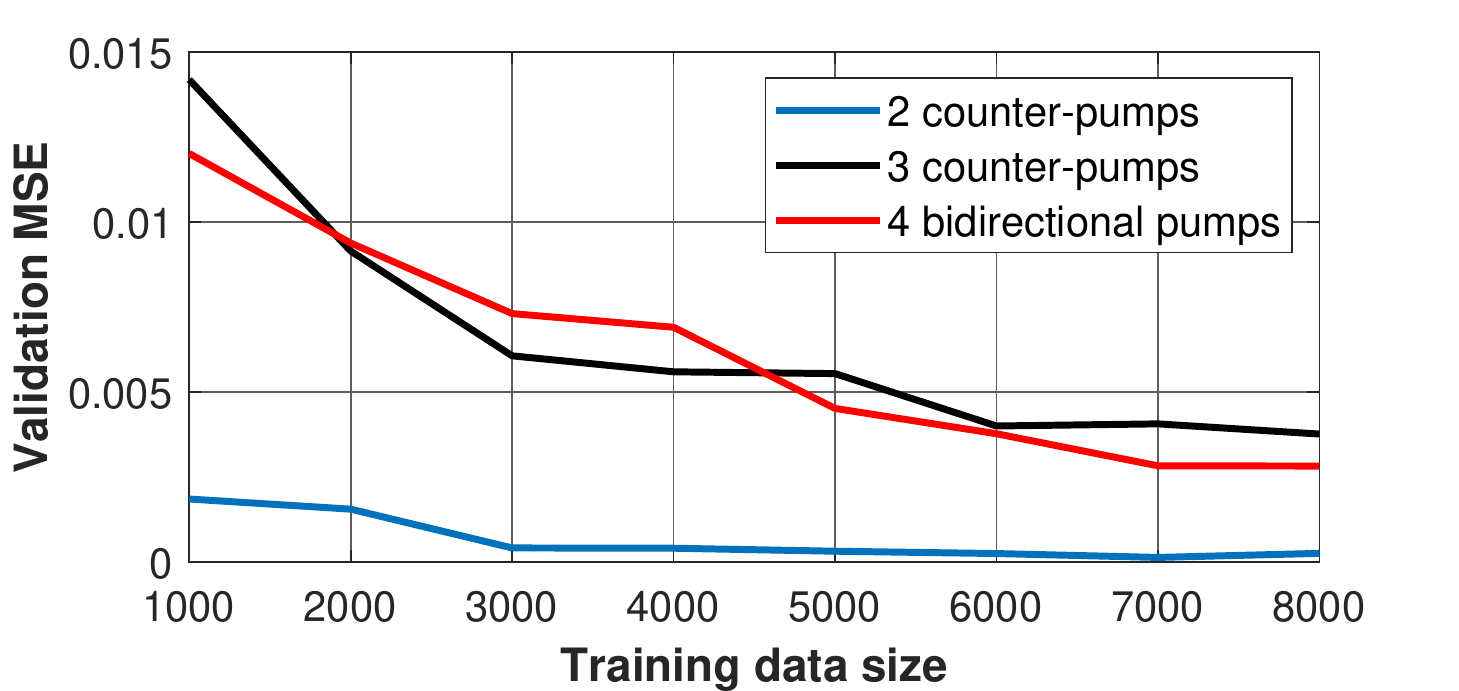}
\caption{MSE of the validation data-set for different pumping schemes as a function of the training data-set size}
\label{fig:Validation}
\end{figure}

In order to determine the size of the training data, for each pumping case, different sizes from 1000 to 8000 have been investigated. Models trained on training data-sets with different size, have been evaluated by the MSE on a validation data-set with 1000 points generated separately. Fig. \ref{fig:Validation} shows the validation MSE as a function of the size of the training data-set. Based on the validation MSE and also the training time, we realized that for 2 and 3 counter and 4 bidirectional cases, best training data sizes are 5000, 6000 and 7000 samples, respectively. Increasing the training size will not result in a remarkable improvement. 

Regarding the parameters of the \textit{feature extraction} network, number of filters $(n_1, n_2, n_3)$, filter sizes $(f_1, f_2, f_3)$, and the average-pooling layer window sizes $(m_1, m_2, m_3)$ have been set and evaluated based on the most common values in the literature. For the number of filters of each layer, we tried 32 and 64 numbers and observed that 64 filters extremely increase the training time with no improvement in performance. Moreover, filter size of $3\times 3$ showed a better validation MSE over a bigger $5\times 5$ filter. We also figured out that for the window size of the average-pooling layers, a commonly used $2\times 2$ window has a better MSE over a window of size $3\times 3$. Furthermore, regarding the regression network parameters, we evaluated the validation MSE by setting the $N_{h1}$ and $N_{h2}$ based on the set $\{20,40,80,100\}$ and realized that for 2 pumping case, $N_{h1}=40$ and $N_{h2}=40$ with ReLU activation, and contrarily, for both 3 and 4 pumping cases, $N_{h1}=100$ and $N_{h2}=40$ with ReLU activation function will minimize the validation loss. For all pumping schemes, the batch size in training phase has been set to 128 and the learning rate of the RMSprop is set to 0.001. Furthermore, the best distance resolution for 2 and 3 pumps is 2 $km$ and for 4 pumps is 1 $km$. Moreover, due to the different value ranges between pump powers and wavelengths at the output layer, the network is trained on the min-max normalized pump configuration vector. The resulting normalized pump configuration vector can be linearly mapped to the desired interval of powers and wavelengths based on the specified ranges for these parameters. 

\begin{figure}
\centering
\includegraphics[width=8.5cm]{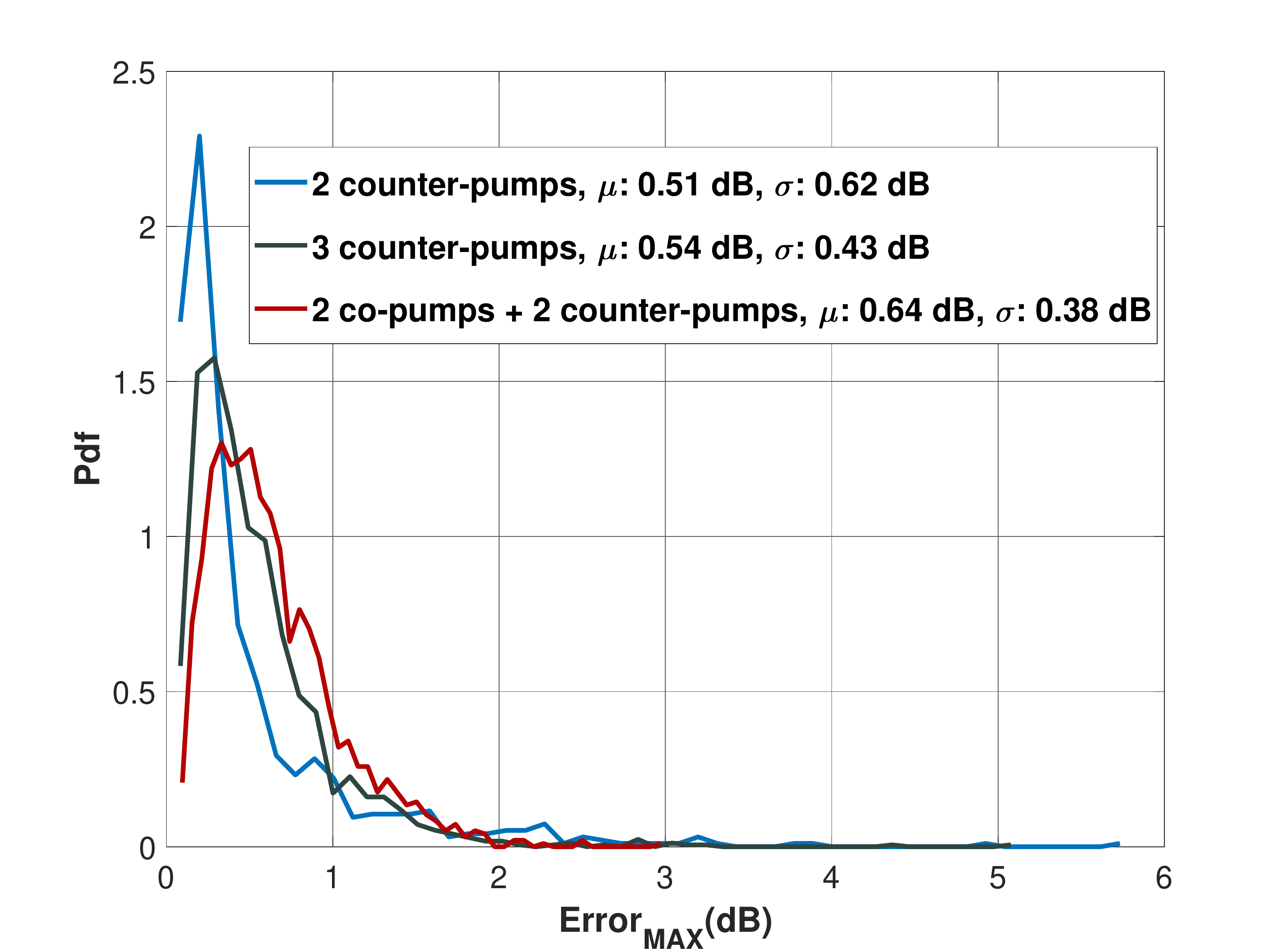}
\caption{Probability density function of the $Error_{max}$. Mean $\mu$ and standard deviation $\sigma$ are shown for 2, 3 and 4 pumps. }
\label{fig:result_fig}
\end{figure}

Final evaluation of the trained models is performed based on the scheme illustrated in Fig.\ref{fig:Evaluation} on test data-sets with 2000 samples generated for each pumping case. Fig.\ref{fig:result_fig} indicates the probability density function (pdf) for the maximum absolute error ($Error_{max}$) of the reconstructed power profile beside its mean $(\mu)$ and standard deviation $(\sigma)$ for all cases. It has been demonstrated that the $\mu$ value for 2 counter-propagating, 3 counter-propagating and 4 bidirectional propagating cases is almost 0.51 $dB$, 0.54 $dB$ and 0.61 $dB$, respectively, and also the $\sigma$ value for these cases is almost 0.62 $ dB$, 0.43 $dB$ and 0.38 $dB$, respectively. We can therefore assert that the proposed method is highly accurate for designing Raman amplifiers based on the signal power profile over a wide band and along the span. 




\section{Conclusion}

A CNN framework is presented for inverse DRA design based on desired signal power profile in frequency and distance domain. The proposed method consists of two networks trained end-to-end: 1) a \textit{feature extraction} with 3 CNN layers employed to extract informative features of the 2D signal power profile and 2) \textit{a regression} aiming to predict the pump powers and wavelengths values based on the extracted features. Numerical simulations show that the proposed framework provide high accuracy in terms of predicting the pump parameters for both counter and bidirectional propagating pumps in C-band.  
\bigskip

\noindent \textbf{Acknowledgments.} This work was financially supported by the European Research Council (ERC-CoG FRECOM grant no. 771878), the Villum Foundation (OPTIC-AI grant no. 29334), and the Italian Ministry for University and Research (PRIN 2017, project FIRST).

\noindent \textbf{Disclosures.} The authors declare no conflicts of interest.

\bibliography{sample}



\end{document}